\documentclass[epj]{svjour}
\usepackage{graphics}

\newcommand{\mc}{\multicolumn}
\newcommand{\gsim}{\mathrel{\mathop{\kern 0pt \rlap
  {\raise.2ex\hbox{$>$}}}
  \lower.9ex\hbox{\kern-.190em $\sim$}}}
\newcommand{\beq}{\begin{equation}}
\newcommand{\eeq}{\end{equation}}

\begin{document}
\title{Weak decay of hypernuclei -- Theoretical status}
\author{Gianni Garbarino
}                     
%
%
\institute{Dipartimento di Fisica Teorica, Universit\`a di Torino and
INFN, Sezione di Torino, I--10125 Torino, Italy}
\date{Received: date / Revised version: date}
%
\abstract{
The physics of the weak decay of hypernuclei is briefly reviewed from a 
theoretical point of view. Special regard is devoted to the recent progress
concerning the determination of the non--mesonic decay widths
and the asymmetry parameters. While 
convincing evidence has been achieved for a solution of the long--standing 
puzzle on the ratio $\Gamma_n/\Gamma_p$, the discrepancies between theory 
and experiment on the decay asymmetries clearly highlight the 
exigence of dedicating further efforts in exploring new aspects of the 
dynamics underlying the non--mesonic weak decay.
%
} 
\maketitle

\section{Introduction}
\label{intro}

Hypernuclei can be considered as a powerful ``laboratory" for
unique investigations of baryon--baryon strangeness--changing
weak interactions. The best studied systems are nuclei containing
a $\Lambda$ hyperon. In such nuclei the $\Lambda$ can decay mesonically,
by emitting a nucleon and a pion, $\Lambda \to \pi N$, 
as it occurs in free space; in addition,
the hyperon (weak) interaction with the nucleons opens new decay channels, 
customarily indicated as non--mesonic modes. In particular,
one can distinguish between one-- and two--nucleon induced
decays, $\Lambda N \to nN$ and $\Lambda NN \to nNN$,
according whether the hyperon interacts with a single nucleon or
with a pair of correlated nucleons.

The field of hypernuclear non--mesonic weak decay
has recently experienced a phase of renewed interest,
thanks to the ideation and accomplishment of innovative experiments as well as to the
advent of elaborated theoretical models. For many years the main open problem 
in the decay of hypernuclei has been the discrepancy between theoretical and
experimental values of the ratio $\Gamma_n/\Gamma_p$ between the neutron--
and proton--induced decay widths, $\Gamma_n\equiv \Gamma(\Lambda n \to nn)$ 
and $\Gamma_p\equiv \Gamma(\Lambda p \to np$). This topic will be discussed
in this contribution together with the most recent evidences 
for a solution of the puzzle.

Another interesting and open question we shall deal with
concerns the asymmetric non--mesonic decay of polarized hypernuclei.
Also in this case, as for the $\Gamma_n/\Gamma_p$ puzzle, we expect
important progress from the present and future improved experiments,
which will provide a guidance for a deeper theoretical understanding of
the hypernuclear decay mechanisms.
                                                                                           
For comprehensive theoretical reviews on hypernuclear weak 
decay we refer the reader to Refs.~\cite{prep,Os98} and references therein.
The experimental viewpoint on the same subject has been discussed
at this conference by H. Outa \cite{OutaHYP}.

\section{Weak decay modes of $\Lambda$ hypernuclei -- General properties}

When a hypernucleus containing a single $\Lambda$ hyperon is stable with respect 
to electromagnetic and strong processes, it is in the ground state, with
the hyperon in the $1s$ level of the $\Lambda$--nucleus mean potential. 
From such a state the hypernucleus then
decays via a strangeness--changing weak interaction, through the
disappearance of the $\Lambda$. 

The mesonic decay mode is the main decay channel of a $\Lambda$ in free space,
including the two channels $\Lambda \to \pi^- p$ and $\Lambda \to \pi^0 n$
with decay rates $\Gamma_{\pi^-}$ and $\Gamma_{\pi^0}$, respectively.
The experimental ratio for the free decay,
$\Gamma^{\rm free}_{\pi^-}/\Gamma^{\rm free}_{\pi^0}\simeq  1.78$,
is very close to 2 and thus strongly suggests the $\Delta I=1/2$ rule
on the isospin change.
As it occurs in the decay of the $\Sigma$ hyperon and in pionic kaon
decays, this rule is based on experimental observations.
Unfortunately, its dynamical origin is not yet 
convincingly understood on theoretical grounds.

The $Q$--value for the mesonic decay
$Q_{\rm M}\simeq m_{\Lambda}-m_N-m_{\pi}\simeq 40$ MeV corresponds
to a momentum of the final nucleon of about $100$ MeV.
As a consequence, in nuclei the $\Lambda$ mesonic decay is disfavored by the
Pauli principle, particularly in medium and heavy systems. It is strictly forbidden in
normal infinite nuclear matter (where the Fermi momentum is
$k_F^0\simeq$ 270 MeV), while in finite nuclei it can occur because of
three important effects: \\
1. In nuclei the hyperon has a momentum distribution
that admits larger momenta for the final nucleon; \\
2. The final pion feels an attraction by the medium such that for fixed
three--momentum $\vec q$ it has an energy smaller than the free one
[$\omega(\vec q)<\sqrt{\vec q\,^2+m_{\pi}^2}$], and consequently, due to
energy conservation, the final nucleon again has more chance to come out above
the Fermi surface; \\
3. At the nuclear surface the local Fermi momentum is considerably
smaller than $k_F^0$, and the Pauli blocking is less effective in
forbidding the decay.

The mesonic channel can provide valuable
information on the pion--nucleus optical
potential since the decay widths $\Gamma_{\pi^-}$ and $\Gamma_{\pi^0}$ 
turn out to be very sensitive to the $\pi^-$ and $\pi^0$ self--energies in the 
nuclear medium \cite{Os93}.
                                                                                           
In hypernuclei the $\Lambda$ decay also occurs through processes which involve
a weak interaction of the hyperon with one or more nucleons.
When the pion emitted by the weak vertex
is virtual, it gets absorbed by the nuclear medium, resulting in
non--mesonic processes of the following type:
\[
\Lambda n  \rightarrow  nn \,\, \left(\Gamma_n\right),\,\,\,
\Lambda p  \rightarrow  np \,\, \left(\Gamma_p\right),\,\,\,
\Lambda NN  \rightarrow  nNN \,\, \left(\Gamma_2\right) . \nonumber
\]
More massive mesons than the pion can also mediate these decays 
\cite{Du96,Pa97}. Hybrid models adopting direct quark mechanisms
in addition to meson--exchange potentials have also been proposed \cite{Sa00}.
In these approaches, the baryon--baryon
short range repulsion originates from quark--exchange between baryons.

The non--mesonic mode is only possible in nuclei and,
nowadays, the systematic study of hypernuclear decay is the only
practical way to get phenomenological information on the
hyperon--nucleon weak interactions.
This is especially facilitated by the fact that    
the final nucleons in the non--mesonic processes have large
momenta ($p_N\simeq 420$ (340) MeV for the one--nucleon
(two--nucleon) induced channel). Therefore, the non--mesonic mode is not
blocked by the Pauli principle; on the contrary, it dominates over the mesonic mode 
for all but the $s$--shell hypernuclei.
Being characterized by a large momentum transfer,
the non--mesonic decay mode is only slightly affected by the
details of hypernuclear structure,
thus providing useful information directly on the hadronic weak interaction.

\begin{figure}
\begin{center}
\resizebox{0.4\textwidth}{!}{%
  \includegraphics{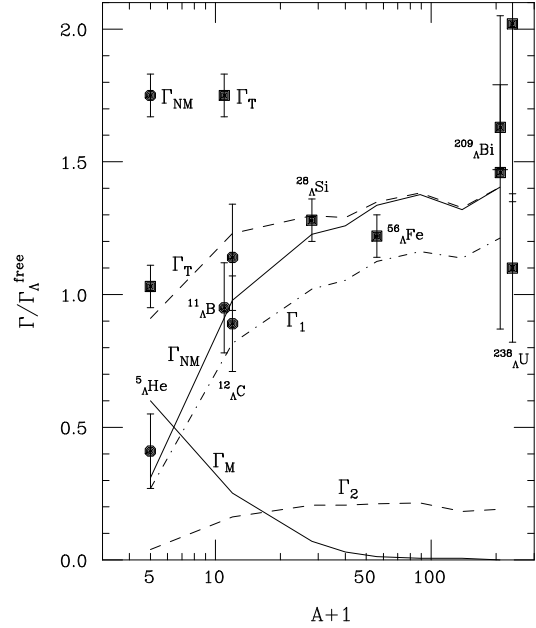}
}
\caption{Decay widths of the $\Lambda$ in finite nuclei as a
function of the nuclear mass number $A$ (taken from Refs.~\cite{prep,Al99}).}
\label{satu}
\end{center}
\end{figure}

The total weak decay rate of a $\Lambda$ hypernucleus is
$\Gamma_{\rm T}=\Gamma_{\rm M}+\Gamma_{\rm NM}$, where
$\Gamma_{\rm M}=\Gamma_{\pi^-}+\Gamma_{\pi^0}$, $\Gamma_{\rm NM}=\Gamma_1+\Gamma_2$
and $\Gamma_1=\Gamma_n+\Gamma_p$, while
the lifetime is $\tau=\hbar/\Gamma_{\rm T}$.
It is interesting to observe that there is an anticorrelation between
mesonic and non--mesonic decay modes such that the total decay rate
is quite stable from light to heavy hypernuclei. This behavior
is evident from Figure~\ref{satu} and is due to
the rapid decrease of the mesonic width and to 
the saturation property of the ${\Lambda}N\rightarrow nN$ interaction
for increasing nuclear mass number.

\section{The ratio $\mathbf{\Gamma_n/\Gamma_p}$}
\label{ratio}

Up to very recent times, the main challenge of hypernuclear weak decay studies
has been to provide a theoretical explanation of the large experimental values 
for the ratio ${\Gamma}_n/{\Gamma}_p$ \cite{prep,Os98}. During this period,
large uncertainties involved in the extraction of
this ratio from data did not allow to reach any definitive
conclusion. These ``old "data \cite{Sz91,No95,No95a}
were quite scarce and not precise due to the difficulty 
of detecting the products of the non--mesonic decays, especially the
neutrons. Moreover, up to now it has not been possible to distinguish between
nucleons produced by the one--body induced and the
(non--negligible) two--body induced decay mechanisms. These limitations
lead to very indirect experimental 
determinations of the decay rates $\Gamma_n$ and $\Gamma_p$
from single--nucleon spectra measurements.
Particularly in the last few years,
this persistent, puzzling status has encouraged a renewed interest for 
hypernuclear non--mesonic decay.
Thanks to recent theoretical \cite{Sa00,Os01,Pa01,It02,Ga03,Ga04,Ba03,Bauer,Bau06}
and experimental \cite{OutaHYP,Ki02,Ok04,OutaVa,Ka06,Kim06} progress, we 
can now safely affirm that the $\Gamma_n/\Gamma_p$ puzzle has been solved. 
A decisive role in this achievement has been played by 
the first measurements of nucleon--coincidence spectra together with
a non--trivial interpretation of data, which required theoretical analyses of two--nucleon 
induced decays and accurate studies of nuclear medium effects on the weak decay 
nucleons. We summarize these important developments in the following.

One--pion--exchange (OPE) models predict small ratios, typically around
0.1-0.2 for the most studied systems, $^5_\Lambda$He and $^{12}_\Lambda$C.
This is mainly due to the particular form of
the OPE potential, which has a strong tensor component,
$\Lambda N(^3S_1)\to nN (^3D_1)$, 
requiring isospin $0$ $np$ pairs in the antisymmetric final state.
On the contrary, the OPE model
has been able to reproduce the total non--mesonic rates measured 
for the above mentioned hypernuclei \cite{prep,Os98}.

Other interaction mechanisms beyond the OPE might then be responsible for the
overestimation of $\Gamma_p$ and the underestimation of $\Gamma_n$.
Many attempts have been made in order to solve the $\Gamma_n/\Gamma_p$ puzzle.
We recall here the inclusion in the ${\Lambda}N\rightarrow nN$
transition potential of mesons heavier than the pion
\cite{Du96,Pa97,Os01,Pa01,It02}, the implementation of interaction terms that
explicitly violate the ${\Delta}I=1/2$ rule
\cite{Pa98,Al99b} and the description of the
short range baryon--baryon interaction in terms of quark degrees of freedom
\cite{Sa00}, which automatically introduces $\Delta I=3/2$ contributions.

A few investigations with transition
potentials including heavy--meson--exchange and/or direct quark (DQ) contributions
have recently improved the situation, without providing
an explanation of the origin of the puzzle.
As discussed in the next paragraphs, the proper determination of $\Gamma_n/\Gamma_p$
from data indeed required an analysis of the two--nucleon induced
decay mechanism and an accurate evaluation of the final state 
interactions  suffered by the detected nucleons.
In Tables~\ref{best-ratio} and \ref{best-nm}
we summarize the results of those calculations
which predicted ratios considerably enhanced with respect to the
OPE values together with experimental data. The variety of 
models adopted in the quoted
works has been extensively discussed in Ref.~\cite{prep}. Here we only 
mention that, with respect to the OPE results, 
the addition of kaon--exchange is found to considerably
reduce $\Gamma_{\rm NM}$ while increasing $\Gamma_n/\Gamma_p$ 
by a factor of about 4. 
This result is mainly due to i) the enhancement of the parity--violating
$\Lambda N(^3S_1)\to nN(^3P_1)$ transition
contributing especially to neutron--induced decays and ii)
the tensor component of kaon--exchange, which has opposite sign
with respect to the OPE one, thus increasing $\Gamma_n$
and reducing $\Gamma_p$. Also the DQ mechanism revealed to be important
to obtain larger $\Gamma_n/\Gamma_p$. All the approaches of Tables~\ref{best-ratio}
and \ref{best-nm} but the one of Ref.~\cite{Bauer}
reproduce the observed non--mesonic widths.
We note that the use of a more realistic
$\Lambda$ wave function would lead to a reduction of about 25\% \cite{Bau06}
of the non--mesonic rate evaluated in Ref.~\cite{Bauer}.
Although no calculation of Table~\ref{best-ratio} is able to explain the 
data of Refs.~\cite{Sz91,No95,No95a,Sa04}, extracted from single--nucleon measurements, 
some predictions are in agreement with the recent determinations of Refs.~\cite{Ga04,Bau06} 
obtained by fitting the nucleon--nucleon coincidence data of 
Refs.~\cite{OutaHYP,OutaVa,Ka06,Kim06}. In the remainder of the present Section 
we discuss these recent achievements, which made possible a solution of
the $\Gamma_n/\Gamma_p$ puzzle.
\begin{table*}
\caption{Theoretical and experimental determinations of the 
$\Gamma_n/\Gamma_p$ ratio.}
\label{best-ratio}
\begin{center}
\begin{tabular}{l c c} \hline
\mc {1}{c}{Ref. and Model} &
\mc {1}{c}{$^5_{\Lambda}$He} &
\mc {1}{c}{$^{12}_{\Lambda}$C} \\ \hline
Sasaki et al. \cite{Sa00} &0.70 &  \\
{\small $\pi +K+$ DQ}               & & \\
Jido et al. \cite{Os01} & &0.53  \\
$\pi +K+2\pi/\sigma+2\pi +\omega$      & & \\
Parre\~{n}o and Ramos \cite{Pa01} &$0.34\div 0.46$ & $0.29\div 0.34$  \\
$\pi +\rho +K + K^* + \omega +\eta$                        & &  \\
Itonaga et al. \cite{It02}    & 0.39 & 0.37  \\
$\pi + 2\pi/\sigma + 2\pi/\rho+\omega$ & &      \\ 
Barbero et al. \cite{Ba03} & 0.24 & 0.21  \\
$\pi +\rho +K + K^* + \omega +\eta$ &  &      \\ 
Bauer and Krmpoti\'c \cite{Bauer} &  & 0.29  \\
$\pi +\rho +K + K^* + \omega +\eta$ &  &      \\ \hline
BNL \cite{Sz91} &$0.93\pm0.55$ 
&$1.33^{+1.12}_{-0.81}$  \\
KEK \cite{No95} & &$1.87^{+0.67}_{-1.16}$  \\
KEK \cite{No95a} &$1.97\pm0.67$ &  \\
KEK--E307 \cite{Sa04} & &$0.87\pm 0.23$  \\
KEK--E462 \cite{Ka06} & $0.45\pm 0.11\pm 0.03$ &  \\
KEK--E508 \cite{Kim06} & & $0.51\pm 0.13\pm 0.05$  \\
KEK--E462/E508 (analysis of Ref.~\cite{Ga04}) 
 & $0.40\pm 0.11$ ($1N$) & $0.38\pm 0.14$ ($1N$)\\
 & $0.27\pm 0.11$ ($1N+2N$) & $0.29\pm 0.14$ ($1N+2N$) \\ 
KEK--E462/E508 (analysis of Ref.~\cite{Bau06})
 &   & $0.37\pm 0.14$ ($1N$)\\
 &   & $0.34\pm 0.15$ ($1N+2N$) \\
\hline
\end{tabular}
\end{center}
\end{table*}
\begin{table*}
\caption{Theoretical and experimental determinations of the
non--mesonic width $\Gamma_{NM}$ (in units of $\Gamma^{\rm free}_{\Lambda}$).
Only the one--nucleon induced decay channel has been taken into account
in the theoretical evaluations.}
\label{best-nm}
\begin{center}
\begin{tabular}{lcc} \hline
\mc {1}{c}{Ref. and Model} &
\mc {1}{c}{$^5_{\Lambda}$He} &
\mc {1}{c}{$^{12}_{\Lambda}$C} \\ \hline
Sasaki et al. \cite{Sa00} &0.52 &  \\
$\pi +K+$ DQ               & &  \\
Jido et al. \cite{Os01} & &0.77 \\
$\pi +K+2\pi/\sigma+2\pi +\omega$      & &  \\
Parre\~{n}o and Ramos \cite{Pa01} &$0.32\div 0.43$ & $0.55\div 0.73$  \\
$\pi +\rho +K + K^* + \omega +\eta$                        & &  \\
Itonaga et al. \cite{It02}    & 0.42 & 1.06  \\
$\pi + 2\pi/\sigma + 2\pi/\rho+\omega$ & &       \\ 
Barbero et al. \cite{Ba03} & 0.69 & 1.17  \\
$\pi +\rho +K + K^* + \omega +\eta$ &  &      \\ 
Bauer and Krmpoti\'c \cite{Bauer}  & & 1.64    \\
$\pi +\rho +K + K^* + \omega +\eta$ &  &      \\ \hline
BNL \cite{Sz91} &$0.41\pm0.14$    &$1.14\pm0.20$ \\
KEK \cite{No95} & &$0.89\pm0.18$ \\
KEK \cite{No95a} &$0.50\pm0.07$ &  \\
KEK--E307 \cite{Sa04} & &$0.828\pm 0.056 \pm 0.066$ \\ 
KEK--E462 \cite{OutaVa} & $0.424\pm 0.024$ & \\
KEK--E508 \cite{OutaVa} &  &$0.940\pm 0.035$ \\
KEK--E462 \cite{OutaHYP} & $0.411\pm 0.023\pm 0.006$ & \\
KEK--E508 \cite{OutaHYP} &  &$0.929\pm 0.027\pm 0.016$ \\ \hline
\end{tabular}
\end{center}
\end{table*}

The authors of Refs.~\cite{Ga03,Ga04} evaluated neutron--neutron and 
neutron--proton energy and angular correlations for $^5_\Lambda$He and $^{12}_\Lambda$C
and analyzed the corresponding data obtained by the experiments KEK--E462 and KEK--E508.
A one--meson--exchange
(OME) model was used for the $\Lambda N\to nN$ transition in a finite nucleus
framework. The two--nucleon induced decay channel $\Lambda np\to nnp$ was taken
into account via the polarization propagator method in the local density approximation
\cite{Al99}, a model applied for the first time 
to hypernuclear decay in Ref.~\cite{Os85}. The intranuclear cascade code
of Ref.~\cite{Ra97} was employed to simulate the nucleon propagation inside
the residual nucleus.
In Table \ref{ome-che} the ratios $N_{nn}/N_{np}$ predicted by 
the OPE model and two OME models
(OMEa and OMEf, using NSC97a and NSC97f potentials, respectively) 
of Refs.~\cite{Ga03,Ga04} are given for the back--to--back kinematics
($\cos \theta_{NN}\leq -0.8$) and nucleon kinetic energies
$T_n, T_p\geq 30$ MeV. The predictions for $\Gamma_n/\Gamma_p$ are also quoted.
The OME results well reproduce the data, thus indicating a ratio
$\Gamma_n/\Gamma_p\simeq 0.3$ for both hypernuclei, in agreement
with some of the pure theoretical determinations of Table~\ref{best-ratio}.
\begin{table}
\begin{center}
\caption{
Predictions of Refs.~\cite{Ga03,Ga04} for the ratio $N_{nn}/N_{np}$
corresponding to an energy thresholds $T^{\rm th}_N$ of $30$ MeV
and to the back--to--back kinematics ($\cos\, \theta_{NN}\leq -0.8$).
The data are from KEK--E462
and KEK--E508 \protect\cite{OutaHYP,OutaVa,Ka06,Kim06}.}
\label{ome-che}
\begin{tabular}{c c c c c} \hline
\mc {1}{c}{} &
\mc {1}{c}{$^5_\Lambda$He} &
\mc {1}{c}{} &
\mc {1}{c}{$^{12}_\Lambda$C} &
\mc {1}{c}{} \\
       & $N_{nn}/N_{np}$ & $\Gamma_n/\Gamma_p$
       & $N_{nn}/N_{np}$ & $\Gamma_n/\Gamma_p$ \\ \hline
OPE     & $0.25$  & $0.09$  & $0.24$ & $0.08$  \\
OMEa    & $0.51$  & $0.34$  & $0.39$ & $0.29$  \\
OMEf    & $0.61$  & $0.46$  & $0.43$ & $0.34$ \\ \hline
EXP  & $0.45\pm 0.11$  &     &  $0.40\pm 0.10$   &  \\ \hline
\end{tabular}
\end{center}
\end{table}

A weak--decay--model independent analysis of the coincidence data 
of Table~\ref{ome-che} has been performed in Ref.~\cite{Ga03}. The
results are given in Table~\ref{best-ratio}
either by neglecting the two--nucleon stimulated decay channel ($1N$) or by adopting
$\Gamma_2/\Gamma_1=0.20$ for $^5_\Lambda {\rm He}$ and 
$\Gamma_2/\Gamma_1=0.25$ for $^{12}_\Lambda {\rm C}$ ($1N+2N$), as predicted in
Ref.~\cite{Al99}.
The $\Gamma_n/\Gamma_p$ values determined in this way are in agreement
with the pure theoretical predictions of Refs.~\cite{Os01,Pa01,It02,Ba03,Bauer} 
(see Table~\ref{best-ratio}) but
are substantially smaller than those obtained experimentally
from single--nucleon spectra analyses \cite{Sz91,No95,No95a,Sa04}. In our opinion,
this result represents a strong evidence for a solution of the long--standing puzzle on 
the $\Gamma_n/\Gamma_p$ ratio. 

This conclusion has been corroborated by a recent
study \cite{Bau06}, analogous to the one of Ref.~\cite{Ga04} but performed 
within a nuclear matter formalism adapted to finite nuclei 
via the local density approximation (see the results in Table~\ref{best-ratio}). 
At variance with Ref.~\cite{Ga04}, a microscopic model more
in line with the functional approach of Ref.~\cite{Al99a} has been
followed for the two--nucleon induced decays, also including the channels
$\Lambda nn\to nnn$ and $\Lambda pp\to npp$ besides the standard mode
$\Lambda np\to nnp$ of the previous phenomenological approach.

Forthcoming coincidence data from KEK, BNL \cite{Gi01}, J--PARC \cite{jparc}
and FINUDA \cite{FI} could be directly compared with the results
of Refs.~\cite{Ga03,Ga04,Bau06}.
This will permit to achieve better determinations of $\Gamma_n/\Gamma_p$
and to establish the first constraints on $\Gamma_2/\Gamma_1$.

\section{The asymmetry puzzle}
\label{asy}

Despite the recent progress just discussed, the reaction mechanism for
the non--mesonic decay does not seem to be fully understood. Indeed,
a new intriguing problem, of more recent origin, is open: it
concerns a strong disagreement between theory and experiment
on the asymmetry of the angular emission of non--mesonic decay protons from polarized
hypernuclei. This asymmetry is due to the interference between parity--violating and
parity--conserving $\Lambda p \to np$ transition amplitudes \cite{Ba90}, while the
widely considered rates $\Gamma_n$ and $\Gamma_p$ are dominated 
by the parity--conserving part of the interaction.
The study of the asymmetric emission of protons from polarized hypernuclei
is thus supposed to provide new constraints on the dynamics of hypernuclear decay.

The intensity of protons emitted in $\Lambda p \to np$ decays
along a direction forming an angle $\Theta$
with the polarization axis is given by (see Ref.~\cite{Ra92} for more details):
\begin{equation}
\label{wdint}
I(\Theta, J)=I_0(J)\left[1+\mathcal{A}(\Theta, J)\right] ,
\end{equation}
$I_0$ being the (isotropic) intensity for an unpolarized hypernucleus.
In the shell model weak--coupling scheme,
the proton asymmetry parameter takes the following form:
\begin{equation}
\label{asymm-a-vec1}
\mathcal{A}(\Theta, J)=p_{\Lambda}(J)\, a_{\Lambda}\, {\rm cos}\, \Theta ,
\end{equation}
$p_{\Lambda}$ being the polarization of the $\Lambda$ spin and
$a_{\Lambda}$ the \emph{intrinsic} $\Lambda$ asymmetry parameter, which is 
expected to be a characteristic of the elementary process $\Lambda p\to np$.

Nucleon final state interactions (FSI) acting after the non--mesonic process
modify the weak decay intensity
(\ref{wdint}). Experimentally, one has access to a proton intensity
$I^{\rm M}$ which is assumed to have the same $\Theta$--dependence
as $I$:
\begin{equation}
\label{wdint2}
I^{\rm M}(\Theta, J)=I^{\rm M}_0(J)\left[1+p_\Lambda(J)\, a^{\rm M}_\Lambda(J)
\cos \Theta\right] .
\end{equation}
The \emph{observable} asymmetry, $a^{\rm M}_\Lambda(J)$, which is expected to depend 
on the hypernucleus, is then determined as:
\begin{equation}
\label{asym-exp}
a^{\rm M}_\Lambda(J)=\frac{1}{p_\Lambda(J)} \frac{I^{\rm M}(0^{\circ},J)-
I^{\rm M}(180^{\circ},J)}{I^{\rm M}(0^{\circ},J)+I^{\rm M}(180^{\circ},J)} .
\end{equation}

While inexplicable inconsistencies appeared between
the first asymmetry experiments of Refs.~\cite{Aj92,Aj00},
as discussed in Ref.~\cite{prep}, very recent
and more accurate data \cite{OutaHYP,OutaVa,Ma05,Ma06}
favor small $a^{\rm M}_\Lambda$ values, compatible with a vanishing value,
for both $s$-- and $p$--shell hypernuclei. On the contrary, theoretical
models based on OME potentials and/or DQ mechanisms predicted rather large and
negative $a_\Lambda$ values (see Table~\ref{other-res}).
It must be noted that, on the contrary, the mentioned models
have been able to account fairly well for the other weak decay observables
($\Gamma_{NM}$ and $\Gamma_n/\Gamma_p$) measured for $s$-- and $p$--shell hypernuclei. 
\begin{table*}
\begin{center}
\caption{Theoretical and experimental determinations of
the asymmetry parameters ($a_\Lambda$ and $a^{\rm M}_\Lambda$, respectively).}
\label{other-res}
\begin{tabular}{l c c} \hline
\mc {1}{c}{Ref. and Model} &
\mc {1}{c}{$^5_\Lambda {{\rm H}}{\rm e}$} &
\mc {1}{c}{$^{12}_\Lambda {{\rm C}}$} \\ \hline
Sasaki et al. \cite{Sa00} &   & \\
$\pi+K+{\rm DQ}$                & $-0.68$     \\
Parre\~no and Ramos \cite{Pa01} &    & \\
$\pi+\rho+K+K^*+\omega+\eta$    & $-0.68$ & $-0.73$    \\
Itonaga et al. \cite{It03}   &    & \\
$\pi+K+2\pi/\sigma+2\pi/\rho+\omega$    & $-0.33$ &   \\
Barbero et al. \cite{Ba05}           &    & \\
$\pi+\rho+K+K^*+\omega+\eta$            & $-0.54$ & $-0.53$   \\ \hline
  KEK--E160 \cite{Aj92}
 &                &  $-0.9\pm0.3$  \\
  KEK--E278  \cite{Aj00}    & $0.24\pm0.22$  &                  \\
  KEK--E462  \cite{Ma05}   & $0.11\pm0.08\pm0.04$  & \\
  KEK--E462  \cite{Ma06}   & $0.07\pm0.08^{+0.08}_{-0.00}$  & \\
  KEK--E508  \cite{Ma05}    &                & $-0.20\pm0.26\pm0.04$  \\
  KEK--E508  \cite{Ma06}    &                & $-0.16\pm0.28^{+0.18}_{-0.00}$  \\ 
\hline
\end{tabular}
\end{center}
\end{table*}

Concerning the above comparison between theory and experiment,
it is important to stress that, while one predicts
$a_\Lambda(^5_\Lambda{{\rm H}{\rm e}})\simeq a_\Lambda(^{12}_\Lambda{{\rm C}})$,
there is no known reason to expect this approximate equality to be valid
for $a^{\rm M}_\Lambda$.
Indeed, the relationship between $I(\Theta,J)$ of Eq.~(\ref{wdint}) and
$I^{\rm M}(\Theta,J)$ of Eq.~(\ref{wdint2}) can be strongly affected by FSI
of the emitted protons, thus preventing 
a direct comparison between $a_\Lambda$ and $a^{\rm M}_\Lambda$.
To overcome this obstacle, an evaluation of the 
FSI effects on the non--mesonic decay of polarized
hypernuclei has been recently performed \cite{Al05} adopting
the same framework of Refs.~\cite{Ga03,Ga04}.

We summarize here some results of this investigation, which is the first one
evaluating $a^{\rm M}_\Lambda$.
The simulated proton intensities turned out to be well fitted by Eq.~(\ref{wdint2});
one can thus evaluate $a^{\rm M}_\Lambda$ through Eq.~(\ref{asym-exp}).
Table~\ref{results-asy} shows OME predictions
for the intrinsic and observable asymmetries.
As a result of the nucleon rescattering in the nucleus,
$|a_\Lambda|\gsim |a^{\rm M}_\Lambda|$ for any
value of the proton kinetic energy threshold: when $T^{\rm th}_p=0$,
$a_\Lambda/a^{\rm M}_\Lambda\simeq 2$ for $^5_\Lambda {{\rm H}}{\rm e}$
and $a_\Lambda/a^{\rm M}_\Lambda\simeq 4$ for
$^{12}_\Lambda {{\rm C}}$; $|a^{\rm M}_\Lambda|$ increases
with $T^{\rm th}_p$ and $a_\Lambda/a^{\rm M}_\Lambda\simeq 1$
for $T^{\rm th}_p=70$ MeV in both cases.
Values of $a^{\rm M}_\Lambda$ rather independent of the hypernucleus
are obtained for $T^{\rm th}_p=30$, $50$ and $70$ MeV.
\begin{table}
\begin{center}
\caption{Results of Ref.~\cite{Al05} for the asymmetries $a_\Lambda$ and
$a^{\rm M}_\Lambda$.}
\label{results-asy}
\begin{tabular}{l c c} \hline
\mc {1}{c}{FSI, $T^{\rm th}_p (\rm MeV)$} &
\mc {1}{c}{$^5_\Lambda{{\rm H}}{\rm e}$} &
\mc {1}{c}{$^{12}_\Lambda{{\rm C}}$} \\ \hline
no FSI, 0          & $a_\Lambda=-0.68$  & $a_\Lambda=-0.73$   \\
with FSI, 0       & $-0.30$  & $-0.16$  \\
with FSI, 30      & $-0.46$  & $-0.37$  \\
with FSI, 50      & $-0.52$  & $-0.51$  \\
with FSI, 70      & $-0.55$  & $-0.65$  \\ \hline
  KEK--E462 \cite{Ma05}   & $0.11\pm 0.08\pm0.04$  &   \\
  KEK--E462 \cite{Ma06}   & $0.07\pm 0.08^{+0.08}_{-0.00}$  &   \\
  KEK--E508 \cite{Ma05}     & & $-0.20\pm 0.26\pm0.04$  \\ 
  KEK--E508 \cite{Ma06}     & & $-0.16\pm 0.28^{+0.18}_{-0.00}$  \\ \hline
\end{tabular}
\end{center}
\end{table}
The data quoted in the table refer to a $T^{\rm th}_p$ of about 30 MeV;
the corresponding predictions of Ref.~\cite{Al05} barely agree with the
$^{12}_\Lambda{{\rm C}}$ datum but are inconsistent with the observation for
$^5_\Lambda{{\rm H}}{\rm e}$.

Recently, an effective field theory approach based on tree--level
pion-- and kaon--exchange and leading--order contact interactions has been applied
to hypernuclear decay \cite{Pa04}. The coefficients of the considered
four--fermion point interaction have been fitted
to reproduce available data for the non--mesonic decay widths
of $^{5}_\Lambda$He, $^{11}_\Lambda$B and $^{12}_\Lambda$C. In this way,
a dominating central, spin--and isospin--independent
contact term has been predicted. Such term
turned out to be particularly important to reproduce a small
and positive value of the intrinsic asymmetry for $^{5}_\Lambda$He,
as indicated by the recent KEK experiments.
In order to improve the comparison with the observed decay asymmetries
in a calculation scheme based on a meson--exchange model,
this result can be interpreted dynamically as the need for the introduction
of a scalar--isoscalar meson--exchange.

Prompted by the work of Ref.~\cite{Pa04}, models based on
OME and/or DQ mechanisms \cite{Sa05,Ba06} have been supplemented with the
exchange of the scalar--isoscalar $\sigma$--meson. Despite 
the rather phenomenological character of these
works (the unknown $\sigma$ weak couplings are fixed
to fit non--mesonic decay data for $^5_\Lambda$He),
they have clearly demonstrated the importance of $\sigma$--exchange
in the non--mesonic decay. More detailed investigations are needed to
establish the precise contribution of the scalar--isoscalar channel.



\subsection{Conclusions}

Experimental and theoretical studies of nucleon coincidence
spectra have recently lead to a solution of the long--standing puzzle on the ratio
$\Gamma_n/\Gamma_p$. Nowadays, values of $\Gamma_n/\Gamma_p$ around 0.3-0.4 are common to
both theoretical and experimental analyses of  $s$-- and $p$--shell hypernuclei. 
An important role in this achievement
has been played by a non--trivial interpretation of data, which required analyses
of two--nucleon induced decays and accurate studies of nuclear medium effects
on the weak decay nucleons. 

Despite this improvement, the reaction mechanism for the hypernuclear non--mesonic weak 
decay does not seem to be understood in detail. 
Indeed, an intriguing problem remains open. It regards an asymmetry of the 
angular emission of non--mesonic weak decay protons from polarized hypernuclei.
Although nucleon FSI turned out to be an important ingredient also
when dealing with this issue, further investigations are required to
solve the problem.

On the theoretical side, recent indications on the
relevance of the scalar--isoscalar channel seem to suggest novel
reaction mechanisms to improve our 
knowledge of the dynamics underlying the non--mesonic decay.
New and improved experiments more clearly establishing
the sign and magnitude of $a^{\rm M}_{\Lambda}$ for $s$-- and
$p$--shell hypernuclei are also necessary.
Future experimental studies of the
inverse reaction $p n\to p\Lambda $ using polarized proton beams
should also be encouraged: this process could give a rich and clean piece of
information on the $\Lambda$--nucleon weak interaction and especially on
the $\Lambda$ polarization observables \cite{Na99}.


\end{document}